\documentclass{article}
\usepackage{arxiv}

\usepackage{cite}
\usepackage{amsmath,amssymb,amsfonts}
\usepackage{algorithmic}
\usepackage{graphicx}
\usepackage{textcomp}
\usepackage{xcolor}
\usepackage{url}

\makeatletter
\let\@ORGmakecaption\@makecaption
\long\def\@makecaption#1#2{\@ORGmakecaption{#1}{#2}\vskip\belowcaptionskip\relax}
\makeatother

\usepackage[nolist,nohyperlinks]{acronym}
\usepackage{cleveref}

\makeatletter
\newcommand*{\org@overidelabel}{}
\let\org@overridelabel\@verridelabel
\@ifpackagelater{acronym}{2015/03/21}{
  \renewcommand*{\@verridelabel}[1]{%
    \@bsphack
    \protected@write\@auxout{}{\string\AC@undonewlabel{#1@cref}}%
    \org@overridelabel{#1}%
    \@esphack
  }%
}{
  \renewcommand*{\@verridelabel}[1]{%
    \@bsphack
    \protected@write\@auxout{}{\string\undonewlabel{#1@cref}}%
    \org@overridelabel{#1}%
    \@esphack
  }%
}
\makeatother
\Crefformat{figure}{#2Fig.~#1#3}
\Crefmultiformat{figure}{Figs.~#2#1#3}{ and~#2#1#3}{, #2#1#3}{ and~#2#1#3}
\Crefformat{table}{#2Tab.~#1#3}
\Crefmultiformat{table}{Tabs.~#2#1#3}{ and~#2#1#3}{, #2#1#3}{ and~#2#1#3}
\Crefformat{equation}{#2Eq.~#1#3}
\Crefmultiformat{equation}{Eqs.~#2#1#3}{ and~#2#1#3}{, #2#1#3}{ and~#2#1#3}
\crefname{lstlisting}{listing}{listings}
\Crefname{lstlisting}{Listing}{Listings}

\usepackage{tikz}
\usepackage{calc}
\usetikzlibrary{arrows,automata,matrix}
\usetikzlibrary{arrows,arrows.meta}
\usetikzlibrary{positioning,arrows}
\usetikzlibrary{shapes,arrows,fit,calc,positioning,automata}
\usepackage{float}
\usetikzlibrary{positioning}
\usetikzlibrary{arrows,arrows.meta}
\usetikzlibrary{decorations.pathreplacing}
\usetikzlibrary{calc}
\usetikzlibrary{arrows,positioning,automata,shadows,fit,shapes}
\usetikzlibrary{trees}
\usepackage{enumitem}
\usepackage{xcolor}
\newcommand*\circled[1]{\tikz[baseline=(char.base)]{%
            \node[shape=circle,fill=black,draw,inner sep=0.1pt] (char) {\footnotesize #1};}}
\usepackage{amssymb}
\usepackage{pifont}
\newcommand{\cmark}{\ding{51}}%
\newcommand{\xmark}{\ding{55}}%
\newcommand{\omark}{\ding{109}}%

\usepackage{textcomp}
\usepackage{xspace}
\let\OldTexttrademark\texttrademark
\renewcommand{\texttrademark}{\OldTexttrademark\xspace}%

\usepackage{color}
\usepackage{listings,chngcntr}
\definecolor{codegreen}{rgb}{0,0.6,0}
\definecolor{codegray}{rgb}{0.5,0.5,0.5}
\definecolor{codepurple}{rgb}{0.58,0,0.82}
\definecolor{backcolour}{rgb}{0.95,0.95,0.92}

\usepackage{balance}

\usepackage[nolist,nohyperlinks]{acronym}
\begin{acronym}
    \acro{ARP}{Address Resolution Protocol}
    \acro{BMBF}{Federal Ministry of Education and Research}
    \acro{CPU}{Central Processing Unit}
    \acro{DHCP}{Dynamic Host Configuration Protocol}
    \acro{DoS}{Denial of Service}
    \acro{DMA}{Direct Memory Access}
    \acro{HAL}{Hardware Abstraction Layer}
    \acro{HMI}{Human Machine Interface}
    \acro{I2C}[I\textsuperscript{2}C]{Inter-Integrated Circuit}
    \acro{ICMP}{Internet Control Message Protocol}
    \acro{ICS}{Industrial Control System}
    \acrodefplural{ICS}{Industrial Control Systems}
    \acro{IDS}{Intrusion Detection System}
    \acro{IIoT}{Industrial Internet of Things}
    \acro{IO}{input/output}
    \acro{IoT}{Internet of Things}
    \acro{IP}{Internet Protocol}
    \acro{LED}{Light-Emitting Diode}
    \acro{LwIP}{Lightweight IP}
    \acro{MCU}{Microcontroller Unit}
    \acro{MitM}{Man in the Middle}
    \acro{OLED}{Organic Light Emitting Diode}
    \acro{PCB}{Printed Circuit Board}
    \acro{PLC}{Programmable Logic Controller}
    \acrodefplural{PLC}{Programmable Logic Controllers}
    \acro{PoC}{Proof of Concept}
    \acro{RAM}{Random-Access Memory}
    \acro{ROM}{Read-only Memory}
    \acro{RTOS}{Real-time Operating System}
    \acro{SCADA}{Supervisory Control and Data Acquisition }
    \acro{SPI}{Serial Peripheral Interface}
    \acro{STM}{STMicroelectronics}
    \acro{SWD}{Serial Wire Debug}
    \acro{TCP}{Transmisson Control Protocoll}
    \acro{USB}{Universal Serial Bus}
\end{acronym}

\usepackage{filecontents}

\newcommand\copyrighttext{%
  \footnotesize \textcopyright 2019 IEEE. 
  Personal use of this material is permitted.
  Permission from IEEE must be obtained for all other uses,
  in any current or future media,
  including reprinting/republishing this material for advertising or promotional purposes,
  creating new collective works, for resale or redistribution to servers or lists,
  or reuse of any copyrighted component of this work in other works.
  DOI: 10.23919/AE.2019.8867032 -- \url{http://dx.doi.org/10.23919/AE.2019.8867032}
}
\newcommand\copyrightnotice{%
\begin{tikzpicture}[remember picture,overlay]
\node[anchor=south,yshift=10pt] at (current page.south) {\fbox{\parbox{\dimexpr\textwidth-\fboxsep-\fboxrule\relax}{\copyrighttext}}};
\end{tikzpicture}%
}

\title{Network Scanning and Mapping for IIoT Edge Node Device Security}

\author{
 Matthias Niedermaier \\
 matthias.niedermaier@hs-augsburg.de \\
 Hochschule Augsburg
 \And
 Florian Fischer \\
 florian.fischer@hs-augsburg.de \\
 Hochschule Augsburg
 \And
 Dominik Merli \\
 dominik.merli@hs-augsburg.de \\
 Hochschule Augsburg
 \And
 Georg Sigl \\
 sigl@tum.de \\
 TU M\"unchen
}

\begin{document}

\maketitle

\begin{abstract}
The amount of connected devices in the industrial environment is growing continuously,
due to the ongoing demands of new features like predictive maintenance.
New business models require more data, collected by \acs{IIoT} edge node sensors based on inexpensive and low performance \acp{MCU}.
A negative side effect of this rise of interconnections is the increased attack surface,
enabled by a larger network with more network services.
Attaching badly documented and cheap devices to industrial networks often without permission of the administrator even further increases the security risk.
A decent method to monitor the network and detect ``unwanted'' devices is network scanning.
Typically, this scanning procedure is executed by a computer or server in each sub-network.
In this paper, we introduce network scanning and mapping as a building block to scan directly from the \ac{IIoT} edge node devices.
This module scans the network in a pseudo-random periodic manner to discover devices and detect changes
in the network structure.
Furthermore, we validate our approach in an industrial testbed to show the feasibility of this approach.
\end{abstract}

\copyrightnotice


\acresetall
\section{Introduction \label{sec:introduction}}
New business models, like predictive maintenance which takes a proactive approach maintaining machinery
and equipment to keep downtime to a minimum, require more data exchange between production
systems and other devices.
Due to this constant digitalization in industrial plants and in the manufacturing industry,
a higher degree of networking is necessary.
Especially low-cost edge node devices like smart sensors are getting attached to systems enabling \ac{IIoT}
applications\cite{7445139}.
The increased interconnections and the amount of devices enlarge the attack surface, leading to new security challenges\cite{7167238}.
To reduce this security risk within the industrial control network, a suitable option is network monitoring.
Thereby, a distinction between active and passive network monitoring is done.
Passive network monitoring is e.g. done by capturing the traffic on a network switch with a monitoring port,
connected to the sub-network.
This variant passively sniffs the traffic within the network,
without additional traffic sent into the investigated network.
The second method is active scanning, which sending additional packets into the observed network.
Furthermore, the scanning procedure requires ``access'' to every single sub-network or a scan node in every separated network.
On the one hand, the implementation is associated with high efforts, which on the other hand results in high costs.

In this paper, we introduce an active network mapping tool,
as a building block for embedded low-cost edge node devices.
This security building block enables probing of devices~(hosts) and services in the network directly from an edge node device,
connected to this network, without the requirement of additional components.
After the edge node is placed into the network and the system is put into operation for the first time,
the edge node scans the network and learns the structure of the network architecture.
This default network ``fingerprint'' is stored locally on the edge and is compared with further scans.
If the network changes during further scans, this indicates an anomaly, which will be reported e.g. to the operator.

The contributions of edge node based network mapping are:
\begin{itemize}
    \item Easy integration with current network state recognition of other participants.
    \item New devices in the network can be found.
    \item New services with open ports will be detected.
    \item Hosts and services which change the status are recognized.
    \item The information of the network scan is only on the ``intelligent'' edge node,
          so that attackers could not exploit this feature.
          Thus, the edge node only has to trust itself and no third parties.
    \item The network scan is done in a pseudo random manner for load balancing
          and that the attacker can not retrace and exploit the scan process.
\end{itemize}

The paper is structured as followed.
\Cref{sec:methodology} explains the methodology of network scanning and mapping.
In \Cref{sec:implementation} the \acs{PoC} implementation is introduced.
To show the feasibility, in \Cref{sec:evaluation} an evaluation is done.
At the end a conclusion is given in \Cref{sec:conclusion}.

\section{Network Mapping on Edge Nodes \label{sec:methodology}}
In a common industrial system at field level,
which is mostly \acs{IP}-based nowadays,
the network structure rarely or never changes.
This means that changes to the network are either caused by maintenance,
malfunctions or an attack.
Independent of what caused these changes in the network environment, the incident must be detected,
because this is a deviation from regular behavior
and e.g. an operator has to decide how to react.

\subsection{Related Work}
There are a lot of different active and passive network scanners available on the market and discussed in research:

One of the best known network scanner is Nmap\cite{lyon2009nmap}.
It offers a wide variety of scan options, as well as various scripts for further analysis.
However, this requires a comparatively high-performance computer compared to an embedded \ac{MCU} used in an \ac{IIoT} edge node.
Additionally, this is usually done by scanning the network from a central point,
which means that either an additional scan node must be placed in each separate subnet,
or only certain subnets can be scanned.

Wedgbury et al.\cite{wedgbury2015automated} gave an overview of passive network scanners for \acp{ICS}.
Furthermore, different \acs{SCADA} network monitoring tools and scanners are compared by Coffey et
al.\cite{coffey2018vulnerability}.
All these specialized \acs{ICS} scanners also have high requirements on system performance and produce a huge amount of data to process.

An Internet-wide search engine (www.censys.io) for \acs{SCADA} devices was introduced by Durumeric et al.\cite{durumeric2015search}.
This is also capable of scanning internal networks with the help of zgrab2\footnote{https://github.com/zmap/zgrab2}.
However, this is a full featured active scanner that does not run on small edge node devices either.

ModScan, a Modbus/TCP enumeration scanner, was introduced by Bristow et al. \cite{bristow2008modscan}.
A specialized vulnerability network scanner for Siemens devices was introduced by Antrobus et al. \cite{antrobus2016simaticscan}, which is based on a modified version of PLCScan\footnote{https://code.google.com/archive/p/plcscan/}.
However, these basic scanners are written in the python scripting language and are again not suitable for \acp{MCU}.

Wang et al.\cite{wang2006survey} and Radhappa et al. \cite{radhappa2018practical}
summarized the current problems and open research questions that exist in wireless sensor networks.
However, intrusion detection is not handled by them, with regard to port scanner and network mapping on low performance edge node devices.

\subsection{Methodology}
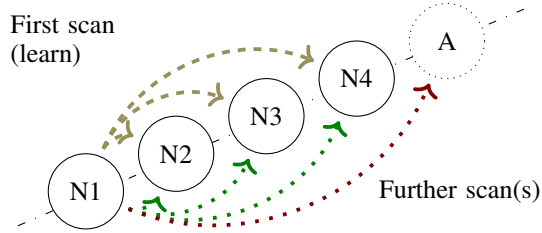
\begin{figure}[htb]
  \centering
\begin{tikzpicture}[node distance=1cm,
    shorten >= 3pt,shorten <= 3pt,
    auto]

     \draw [-, color=black, loosely dashdotted] (-1,-0.5)
      to node[below, text width=1.5cm, text centered]{} (6,2.5);

  \foreach \x in {0,...,3}
    {\pgfmathtruncatemacro{\label}{\x +1}
    \node [circle, draw, fill=white, text width=0.5cm, minimum size=1.0cm,
      inner ysep=0.1cm, text centered] at (\x*1.2,\x*0.5) (n\label) {N\label};}

  \node [circle, draw, fill=white, text width=0.5cm, minimum size=1.0cm, dotted,
      inner ysep=0.1cm, text centered] at (4.8,2) (n5) {A};

  \foreach \x in {2,...,4}
    \draw [->, bend angle=45, bend left, color=yellow!50!black, dashed, line width=0.5mm] (n1)
      to node[below, text width=1.5cm, text centered]{} (n\x);
  \foreach \x in {2,...,4}
    \draw [->, bend angle=45, bend right, color=green!50!black, loosely dotted, line width=0.5mm] (n1)
      to node[below, text width=1.5cm, text centered]{} (n\x);
  \draw [->, bend angle=45, bend right, color=red!50!black, loosely dotted, line width=0.5mm] (n1)
      to node[below, text width=1.5cm, text centered]{} (n5);

  \node [rectangle, text width=2.0cm, minimum size=1.5cm,
      inner ysep=0.1cm] at (0,2) (x) {First scan (learn)};
   \node [rectangle, text width=2.2cm, minimum size=2.2cm,
      inner ysep=0.1cm] at (5,0) (x) {Further scan(s)};

      \end{tikzpicture}
\caption{View of the network mapping from an \ac{IIoT} edge node device of two different scans.}
\label{fig:networkview}
\end{figure}

In this paper, we introduce an edge node based solution in which every smart sensor scans its own network environment.
\Cref{fig:networkview} shows an example of network mapping seen from an edge node device.
In this example network, there are 4 nodes (N1-N4) and N1 is scanning the network in a pseudo random periodic manner.
The dashed arrows show the first scan results with \textcolor{yellow!50!black}{4 devices (N1-N4)} found.
This first scan is used as a reference and it must be ensured that this network is not already contaminated.
After this an additional edge node device gets into the network (A).
Therefore, the second scan (dotted), on one hand detects the \textcolor{green!50!black}{expected devices (N1-N4)} and on the other hand also the \textcolor{red!50!black}{new device (A)}.
This could indicate an intruder or other processes causing a change in the network, e.g. maintenance work.

For the network scanning, followed by the mapping, where the connections of the hosts get analyzed, the following parameters can be used.
These are grouped into two classes which are introduced in the following section, the standard port scan methods and additional options:
\begin{itemize}
    \item \textbf{Host alive discovery}:With an \ac{ICMP} ping sweep, the active hosts in the network can be detected.
    \item \textbf{SYN/connect scan}: Open ports and services are detected with SYN and connect scanning.
    \item Optional: The \textbf{ping timing} can be used to detect redirection, e.g. \ac{MitM} attacks.
    \item Optional: On the application level, e.g. Modbus/TCP, a more detailed \textbf{fingerprint} is possible.
\end{itemize}

Our approach combines the methods presented here and integrates them as a security building block for \ac{IIoT} edge node devices.
To the best knowledge of the authors, no network scanner for low performance \ac{MCU} devices is available at present.

One of the biggest advantages of distributed scan is, that different paths and subnets can be easily scanned.
This is for example the case, if networks are strongly segmented, that the already existing \ac{IIoT} edge devices in this subnets take over the scanning themselves and no additional scanning hardware is necessary, enabled by a software update.

\subsection{Host Alive Discovery}
The first step that is executed in a network scan is the detection of whether a device is active or
not.
This method is called ``host alive discovery'' and is done via a ping sweep on the \ac{IP} \ac{ICMP} level.
If this is done in a local network, this results in an \ac{ARP} request, and if the host answers (\ac{ARP} response), it is up and is tried to be pinged.
If a host does not respond to \ac{ICMP} ping messages,
that does not necessarily mean that it is non-existent,
it may also be possible that it has just disabled \ac{ICMP} echo.
In this case it is possible to do a port scan anyway, which is time consuming.

\subsection{SYN/Connect Scan}
\Cref{{fig:scantype}} shows the flowchart of \textit{SYN} and \textit{connect} scanning.
If the port is closed but the host is up, the target responds with a \textit{RST} directly after the SYN packet.
If the host is up, but does not send any packet at all, then a packet filter is active.
After the target host has answered, a \textit{RST} packet is send if \textit{SYN} scanning is used.
In contrast, if a connect scan is executed the \textit{SYN}+\textit{ACK} message from the target host is acknowledged
and it is possible to get the \textit{Data}/\textit{Banner} from the target.

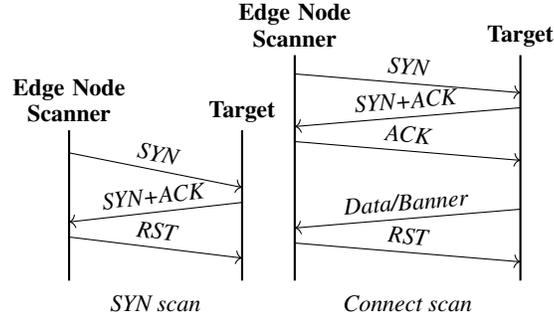
\begin{figure}[htb]
	\centering
	\begin{small}
	\begin{tikzpicture}[node distance=1cm,
	auto
	]
\coordinate (A) at (0,2);
\coordinate (B) at (0,0);
\coordinate (C) at (2.3,2);
\coordinate (D) at (2.3,0);
\draw[thick] (A)--(B) (C)--(D);
\draw (A) node[above, text width=2cm, align=center]{\textbf{Edge Node \\ Scanner} };
\draw (C) node[above]{\textbf{Target} };

\coordinate (x) at ([xshift=0.0cm, yshift=-0.3cm]$(B)!0.5!(D)$);
\draw (x) node[]{\textit{SYN scan}};

\coordinate (E) at ($(A)!0.1!(B)$);
\draw (E) node[left]{\textit{}};

\coordinate (F) at ($(C)!0.43!(D)$);
\draw (F) node[right]{\textit{}};
\draw[->] ([xshift=0.0cm, yshift=-0.1cm]E) -- ([xshift=0.0cm, yshift=0.1cm]F) node[midway,sloped,above]{\textit{\small SYN}};

\coordinate (G) at ($(A)!0.66!(B)$);
\draw (G) node[left]{\textit{}};
\draw[->] ([xshift=0.0cm, yshift=-0.1cm]F) -- ([xshift=0.0cm, yshift=0.1cm]G) node[midway,sloped,above]{\textit{\small SYN+ACK}};

\coordinate (H) at ($(C)!0.9!(D)$);
\draw (H) node[right]{\textit{}};
\draw[->] ([xshift=0.0cm, yshift=-0.1cm]G) -- ([xshift=0.0cm, yshift=0.1cm]H) node[midway,sloped,above]{\textit{\small RST}};

\coordinate (A) at (3,3);
\coordinate (B) at (3,0);
\coordinate (C) at (6,3);
\coordinate (D) at (6,0);
\draw[thick] (A)--(B) (C)--(D);
\draw (A) node[above, text width=2cm, align=center]{\textbf{Edge Node \\ Scanner} };
\draw (C) node[above]{\textbf{Target} };

\coordinate (x) at ([xshift=0.0cm, yshift=-0.3cm]$(B)!0.5!(D)$);
\draw (x) node[]{\textit{Connect scan}};

\coordinate (E) at ($(A)!0.05!(B)$);
\draw (E) node[left]{\textit{}};

\coordinate (F) at ($(C)!0.20!(D)$);
\draw (F) node[right]{\textit{}};
\draw[->] ([xshift=0.0cm, yshift=-0.1cm]E) -- ([xshift=0.0cm, yshift=0.1cm]F) node[midway,sloped,above]{\textit{\small SYN}};

\coordinate (G) at ($(A)!0.35!(B)$);
\draw (G) node[left]{\textit{}};
\draw[->] ([xshift=0.0cm, yshift=-0.1cm]F) -- ([xshift=0.0cm, yshift=0.1cm]G) node[midway,sloped,above]{\textit{\small SYN+ACK}};

\coordinate (H) at ($(C)!0.50!(D)$);
\draw (H) node[right]{\textit{}};
\draw[->] ([xshift=0.0cm, yshift=-0.1cm]G) -- ([xshift=0.0cm, yshift=0.1cm]H) node[midway,sloped,above]{\textit{\small ACK}};

\coordinate (I) at ($(C)!0.65!(D)$);
\draw (I) node[right]{\textit{}};

\coordinate (J) at ($(A)!0.80!(B)$);
\draw (J) node[left]{\textit{}};
\draw[->] ([xshift=0.0cm, yshift=-0.1cm]I) -- ([xshift=0.0cm, yshift=0.1cm]J) node[midway,sloped,above]{\textit{\small Data/Banner}};

\coordinate (K) at ($(C)!0.95!(D)$);
\draw (K) node[right]{\textit{}};
\draw[->] ([xshift=0.0cm, yshift=-0.1cm]J) -- ([xshift=0.0cm, yshift=0.1cm]K) node[midway,sloped,above]{\textit{\small RST}};
	\end{tikzpicture}
	\end{small}
	\caption{Flowchart of a SYN and a connect scan with an open port on the target.}
	\label{fig:scantype}
\end{figure}

The advantage of \textit{SYN} scanning is that the data does not reach the application level, and therefore, there are no log entries in the application.
However, since this is not used as preparation of an attack and it is not the goal to stay under the radar the preferred method is a connect scan, because more information from the target host could be collected.
In addition, Soulie \cite{soullie} recommends to perform connect scans within \ac{ICS} networks to reduce influences on the process.
This is the preferred scanning method especially in fragile \ac{ICS} networks.

\subsection{Pseudo Random Scanning \label{subsec:pseudo_random_scanning}}
On one hand, the selection of the target host to be scanned must be chosen randomly,
as an attacker might otherwise hide himself.
And on the other hand, if more scanners are in the network, to not flood one target.
Furthermore, the start time of the scan is randomly chosen between one and five minutes after the edge node is switched on.
This delay after the start-up is necessary that other devices have finished booting
and if there are multiple scanners in a network the network load will be further distributed.
For attackers the pseudo random scanning on distributed edge nodes makes it difficult to guess scan pattern.

\subsection{Intrusion Detection Handling}
If a new device is detected in the network,
a known device is no longer reachable,
or ports/services have changed,
this should be regarded as an incident.
In this case, the edge node could go into a safe state
or report the incident to a centralized logger.
This depends on the respective field of application.
The advantage here is that the direct processing on the edge node allows a fast
and independent reaction, since there are no dependencies and long run-times,
e.g. through network communication to a central server.

\section{\acs{PoC} Implementation \label{sec:implementation}}
To prove that the approach is feasible on an embedded \ac{MCU},
it was implemented for the usage in a test environment.

\subsection{Hardware}
On the hardware side, the edge node \ac{PoC} consists of a \ac{MCU} from the ARM\textsuperscript{\tiny{\textregistered}} Cortex\textsuperscript{\tiny{\textregistered}}-M7 series (STM32F767),
with a custom \acs{PCB} for \acs{IO} operations.
\Cref{tab_boardinfo} lists the features of the development board with the \ac{MCU}.
The ARM\textsuperscript{\tiny{\textregistered}}Cortex\textsuperscript{\tiny{\textregistered}}-M7 \acp{MCU} are the high performance series of the energy-efficient Cortex\textsuperscript{\tiny{\textregistered}}-M product range.

\begin{table}[htb]
    \centering
    \caption{Specification of the used edge node hardware for the \ac{PoC} implementation.}
    \label{tab_boardinfo}
        \begin{tabular}{l l}
            \hline
            \textbf{Hardware} & \textbf{\ac{IIoT} Edge Node}     \\
            \hline
            Board design      & STMicroelectronics     \\
            \acs{MCU}         & STM32F767ZIT6          \\
            Core              &
            ARM\textsuperscript{\tiny{\textregistered}} Cortex\textsuperscript{\tiny{\textregistered}}-M7 \\
            Clock             & up to 216 MHz          \\
            \acs{RAM}         & 512kB                  \\
            Flash             & 2MB                    \\
        \end{tabular}
\end{table}

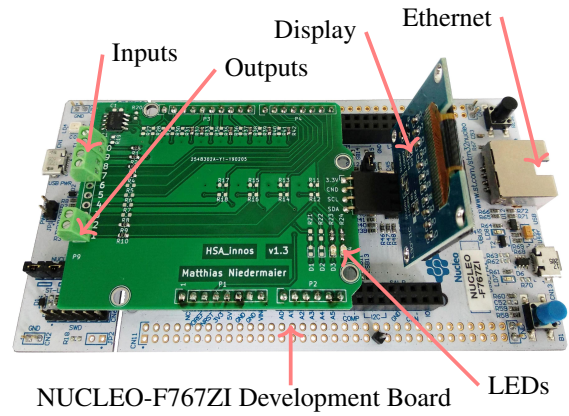
\begin{figure}[htb]
	\centering
	    \begin{tikzpicture}	
        \node[inner sep=0pt] (shield) at (0,0)
            {\includegraphics[width=0.45\columnwidth]{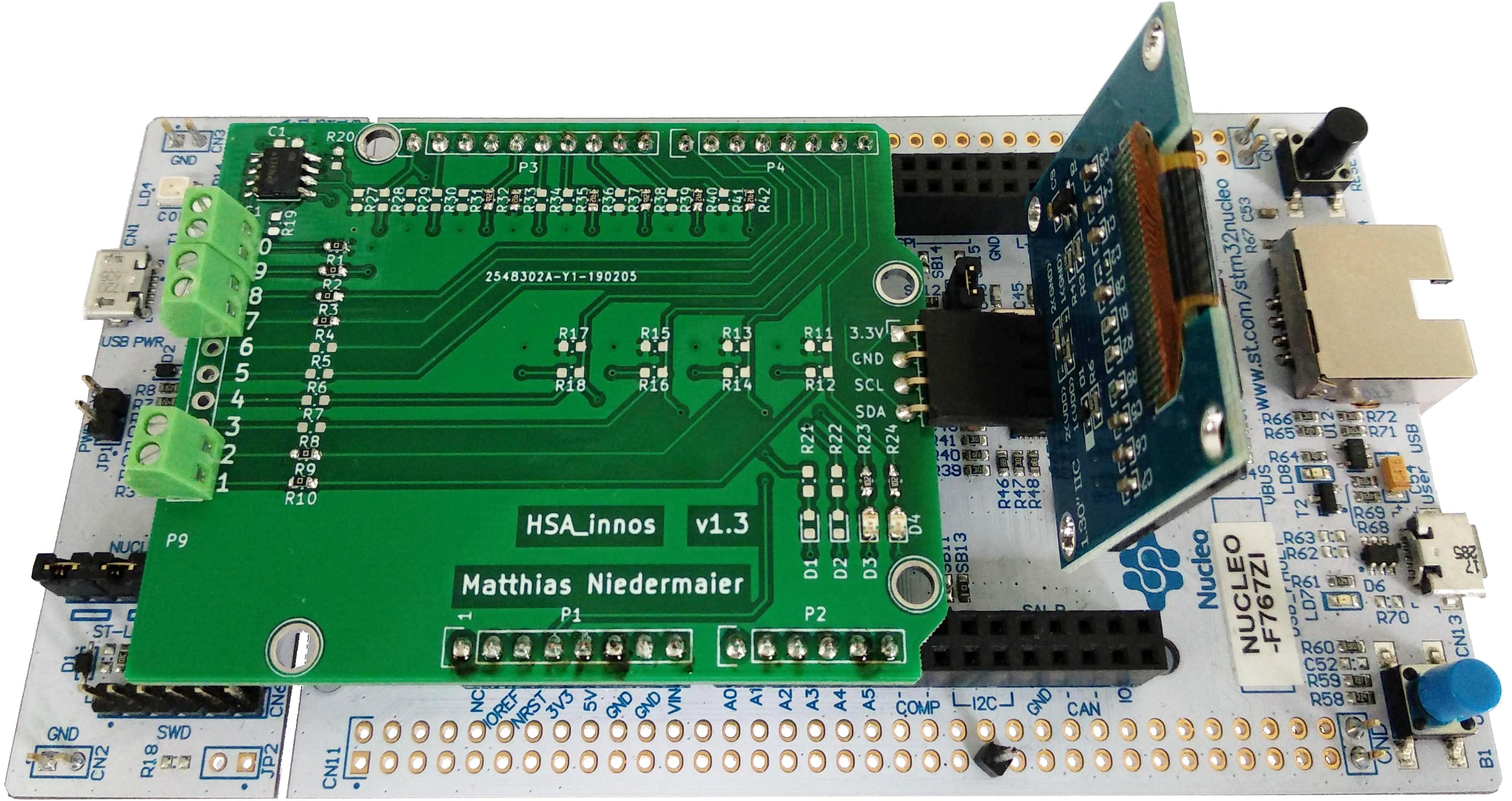}};
        \node[align=left,anchor=east] at (1.0,2.3) {Display};
        \draw[->,color=red!50, line width=0.5mm] (1.0,2.3) -- (1.6,0.7);
        \node[align=left,anchor=east] at (2.8,2.5) {Ethernet};
        \draw[->,color=red!50, line width=0.5mm] (2.8,2.5) -- (3.3,0.5);
        \node[align=left,anchor=west] at (-2.5,2) {Inputs};
        \draw[->,color=red!50, line width=0.5mm] (-2.5,2) -- (-2.7,0.7);
        \node[align=left,anchor=west] at (-1.0,1.8) {Outputs};
        \draw[->,color=red!50, line width=0.5mm] (-1.0,1.8) -- (-2.8,-0.4);
        \node[align=left,anchor=west] at (-3.5,-2.6) {NUCLEO-F767ZI Development Board};
        \draw[->,color=red!50, line width=0.5mm] (0.0,-2.4) -- (0.0,-1.6);
        \node[align=left,anchor=west] at (2.5,-2.4) {LEDs};
        \draw[->,color=red!50, line width=0.5mm] (2.5,-2.4) -- (0.7,-0.6);
    \end{tikzpicture}
	\caption{Hardware platform of each \ac{IIoT} edge node device based on a STMicroelectronics development board with a custom \acs{PCB}.}
	\label{fig:pcb_sensorscan}
\end{figure}

\Cref{fig:pcb_sensorscan} shows the STM development board with our custom \ac{PCB}.
On the right side, the development board has a RJ-45 Ethernet connector
to connect the \ac{IIoT} edge device to the network.
The development board provides an Arduino\texttrademark Uno V3 connector,
where additional shields can be mounted.
This is used for our custom \acs{PCB}, enabling the usage with different \ac{MCU} prototyping boards.
The custom \ac{PCB} provides input and output capabilities and an \acs{I2C} display connector to show current scan information.
Additionally, \acp{LED} are mounted to indicate suspicious behavior.
This can be used as in a data center, where servers which require maintenance flash an \ac{LED}.
This allows a quick finding when many devices are installed in a plant.

\begin{figure}[htb]
	\centering
	    \begin{tikzpicture}
	\node[inner sep=0pt] (shield) at (0,0)
            {\includegraphics[width=0.20\columnwidth]{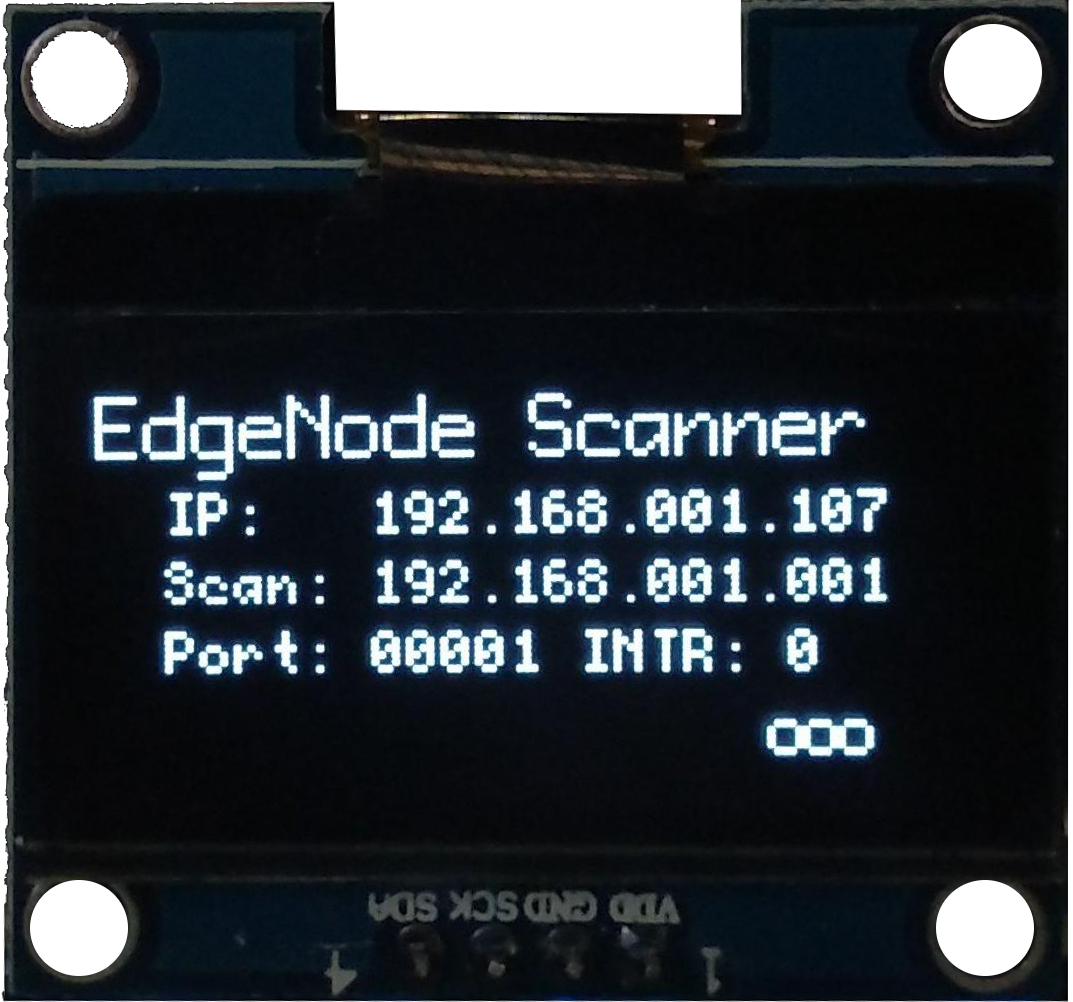}};
    \end{tikzpicture}
	\caption{Scan progress and results on the edge node display.}
	\label{fig:scan_display}
\end{figure}

The current scan progress and intrusion message can be displayed on the display of the edge node device itself.
\Cref{fig:scan_display} shows the 1.3 inch \acs{OLED} display with a SH1106 \ac{I2C} driver.

\subsection{Software}

The software on the \ac{MCU} uses FreeRTOS\footnote{https://www.freertos.org/}
with the LwIP\footnote{https://savannah.nongnu.org/projects/lwip/} stack.
The edge node device provides a Modbus/TCP slave, to control the \acp{IO} and
a web server to provide current information.

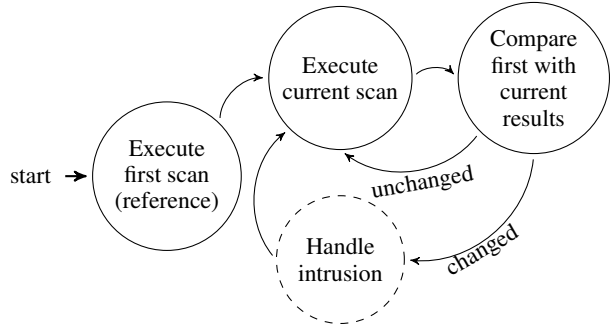
\begin{figure}[htb]
    \centering
    \footnotesize
    \begin{tikzpicture}[>=stealth',
        shorten >=2pt,
        auto,
        shorten <=2pt,
        node distance=0.6 cm,
        every state/.style={align=center,
                            minimum size=1.7cm},
        every edge/.style={draw,thick},
        loop label/.style={draw,align=center,
                           text width=2cm,
                           outer sep=4pt,
                           minimum height=1cm}
        ]

        \node[initial,state] (A)                {Execute \\ first scan \\(reference)};
        \node[state]         (B) [above right= of A, xshift=0.5cm, yshift=-0.5cm]  {Execute \\ current scan};
        \node[state]         (C) [right= of B]  {Compare \\ first with \\ current \\ results};
        \node[state,dashed]  (D) [below= of B]  {Handle \\ intrusion};

        \node[rotate=10] at (3.4,-0.1) {unchanged};
        \node[rotate=30] at (4.2,-1.0) {changed};

        \draw[->, bend angle=45, bend left] ([xshift=0ex]A.north east) to ([xshift=0ex]B.west);
        \draw[->, bend angle=45, bend left] ([xshift=0ex]B.east) to ([xshift=0ex]C.west);
        \draw[->, bend angle=45, bend left] ([xshift=0ex]C.south) to ([xshift=0ex]D.east);

        \draw[->, bend angle=45, bend left] ([xshift=0ex]C.south west) to ([xshift=0ex]B.south);
        \draw[->, bend angle=45, bend left] ([xshift=0ex]D.west) to ([xshift=0ex]B.south west);

    \end{tikzpicture}
    \caption{High level view of the scan process. After the first/trusted scan a continuous monitoring is running.}
    \label{fig:scanprocess}
\end{figure}

The scan process is shown in \Cref{fig:scanprocess}.
The first scan of the network is regarded as a secure state and will be used as a reference for later scans.
After this, the scans are executed periodically and the results are compared with the results of the initial scan, which is treated as a trusted dataset.
In case any mismatch is detected during the following scans, the intrusion handling is initiated.

\vspace*{-0.4cm}

The scan module is designed as a task in FreeRTOS and could be used as a building block in other devices as well.

\begin{figure}[htb]
	\centering
	\includegraphics[width=0.40\columnwidth]{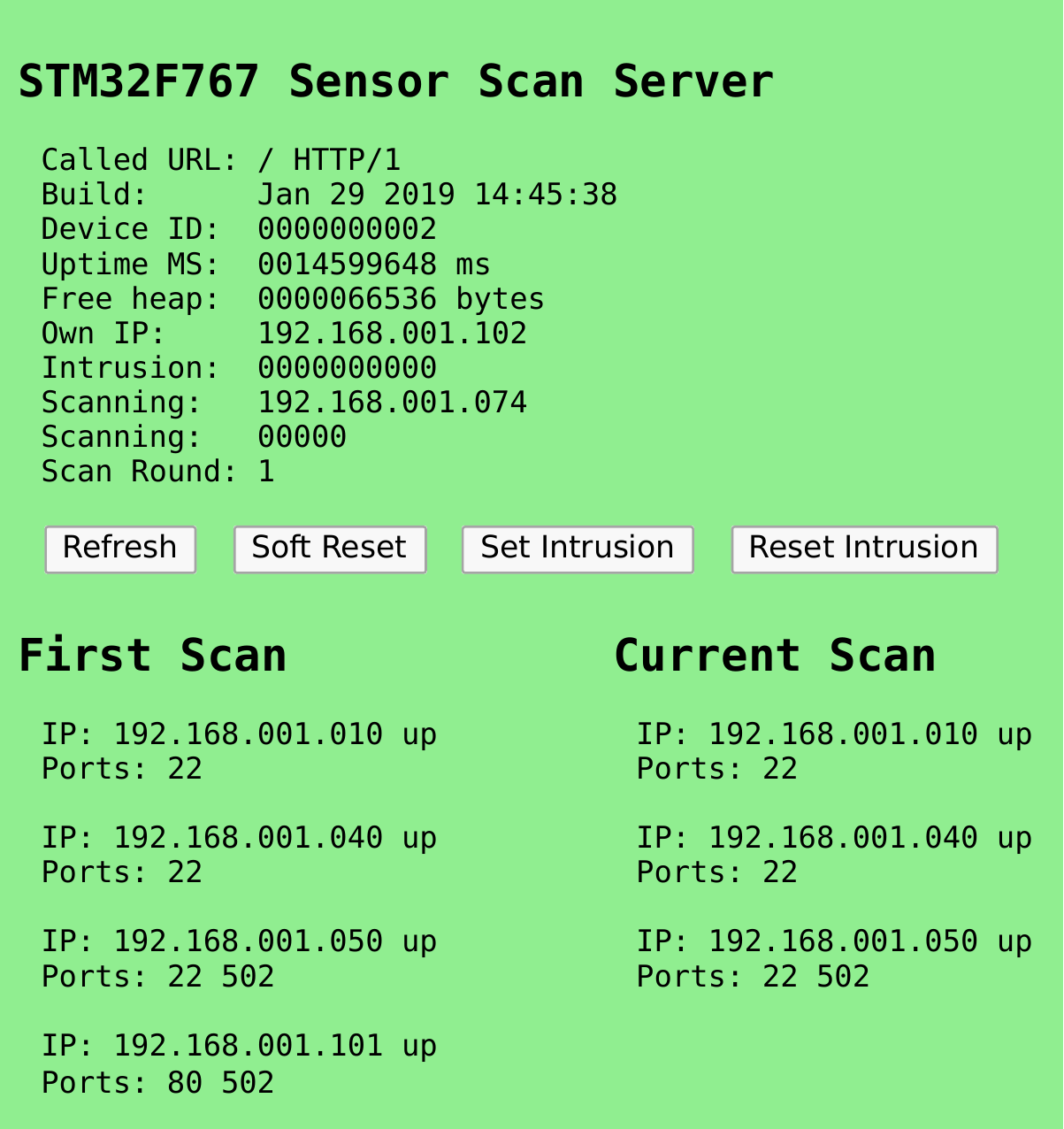}
	\caption{Webpage running on the edge node, displaying the current scan status and debug output.}
	\label{fig:scan_output}
\end{figure}

\Cref{fig:scan_output} shows the current debug output of a scanning edge node device.
On the lower left side, the data of the trusted scan can be seen, which serves as a reference data-set.
On the lower right side, the output of the current scan progress is illustrated.
The \ac{IP} with the alive status is printed and also the open \ac{TCP} ports.

This representation is not for productive usage,
because attackers could use this information for aimed attacks.
Preferably, only the intrusion message with the changes is sent cryptographically protected to a
centralized logger or only allows authenticated user access. 
However, this set-up depends heavily on the integration of the scanners,
e.g. if there are local operators with access to \acp{HMI} 
or a centralized control, who can react to the incident.

\section{Evaluation \label{sec:evaluation}}
To show the feasibility of our approach an evaluation is done.
This is divided into four parts.
First, the feasibility in our open-source industrial testbed is measured,
then the network performance is evaluated,
after this the \ac{MCU} requirements are measured,
and at the end the attack detection is evaluated.

\subsection{Industrial Testbed}
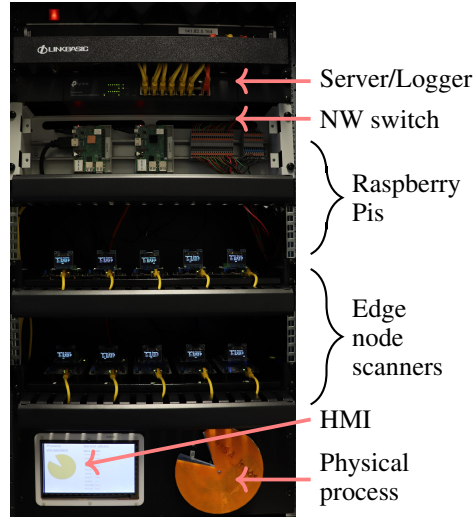
\begin{figure}[htb]
\center
\begin{tikzpicture}
	\node[inner sep=0pt, anchor=south west] (x) at (0,0)
	    {\includegraphics[width=.23\columnwidth]{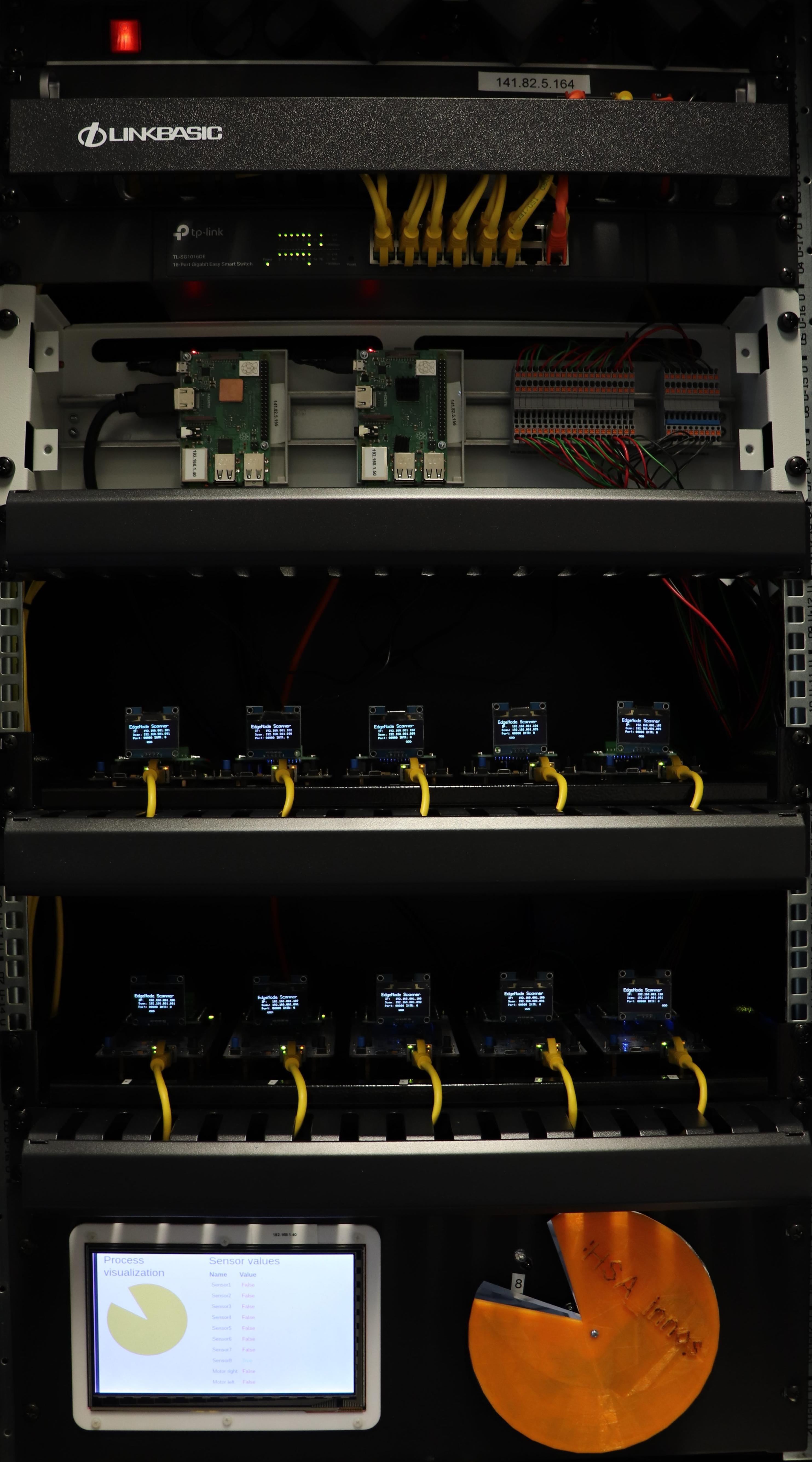}};
\draw[thick,black,decorate,decoration={brace,amplitude=12pt,mirror}] (4.0,1.5) -- (4.0,3.3) node[midway, right,xshift=12pt,text width=1cm]{Edge node \\ scanners};   

\draw[thick,black,decorate,decoration={brace,amplitude=12pt,mirror}] (4.0,3.5) -- (4.0,5.0) node[midway, right,xshift=12pt,text width=1cm]{Raspberry Pis};    

\node[align=right,anchor=west] at (4.0,1.3) {HMI};
\draw[->,color=red!50, line width=0.5mm] (4.0,1.3) -- (1.0,0.7);

\node[align=right,anchor=west, text width=1cm] at (4.0,0.5) {Physical \\ process};
\draw[->,color=red!50, line width=0.5mm] (4.0,0.5) -- (3.0,0.5);

\node[align=right,anchor=west] at (4.0,5.3) {NW switch};
\draw[->,color=red!50, line width=0.5mm] (4.0,5.3) -- (3.0,5.3);

\node[align=right,anchor=west] at (4.0,5.8) {Server/Logger};
\draw[->,color=red!50, line width=0.5mm] (4.0,5.8) -- (3.0,5.8);
\end{tikzpicture}
\caption{Pictures of the open source \ac{ICS} testbed, which is controlling a physical process over Modbus/TCP.}
\label{fig:racksetup}
\end{figure}

The \ac{PoC} implementation is evaluated in our open source industrial testbed (\Cref{fig:racksetup}).
There, the introduced network scanner is running on each \ac{IIoT} edge node,
which are all accessed and controlled by an OpenPLC\cite{alves2014openplc}
instance over Modbus/TCP running on a Raspberry Pi.
Eight edge nodes are each connected to one sensor and one edge node is connected to a motor rotating a disc.
Furthermore, there is a \acs{HMI} displaying the current state of all edge node devices.

\subsection{Network Performance Measurement}
\begin{figure}[htb]
	\centering
	\includegraphics[width=0.99\textwidth]{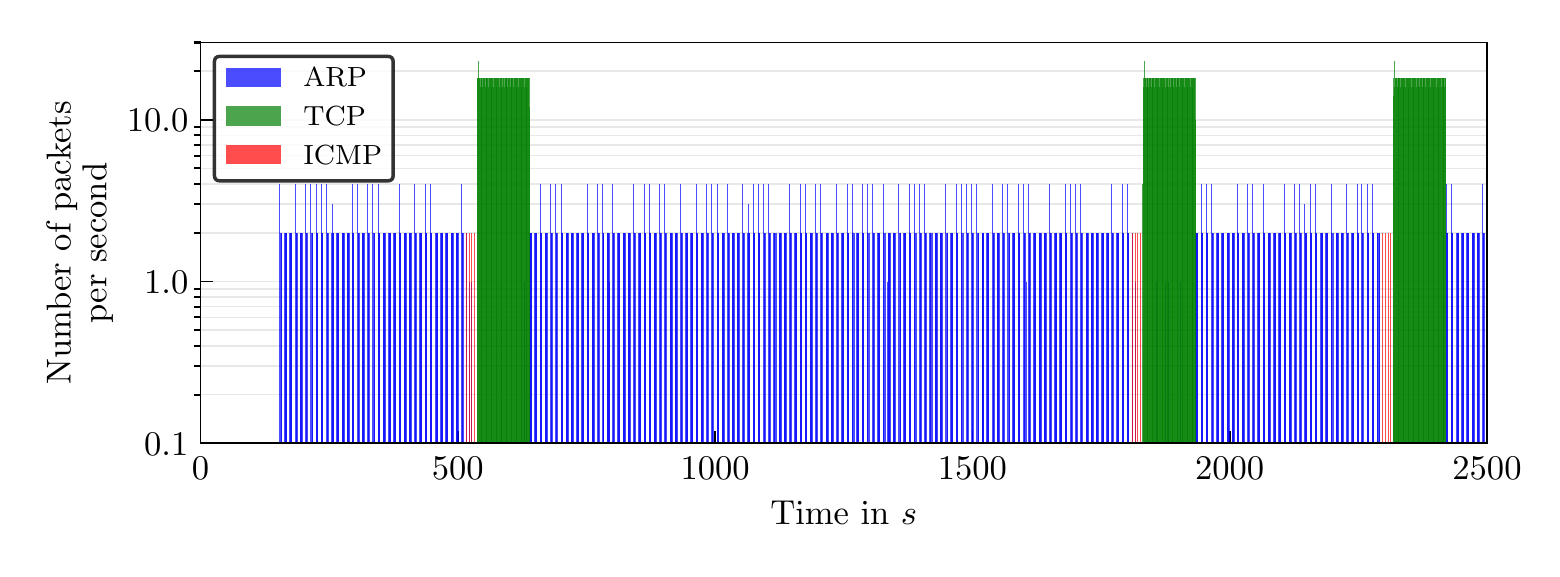}
	\caption{Plot over time, with packets per second of an edge node scanning the network. \acs{ARP} and \acs{ICMP} ping requests are used to check if the hosts are up. \acs{TCP} connect scans are done if the host is up.}
	\label{fig:network_plot}
\end{figure}

\Cref{fig:network_plot} shows the number of packets per second during scanning of one edge node in the open-source testbed.
As shown in \cite{niedermaier18woot}, high scan rates can affect the control behavior of \acp{PLC}.
Therefore, the amount of packets must be low, depending on the components in the network.
As an example, the parameter in our testbed is set to 100$ms$ delay between pings,
which affects the number of packets per second of \ac{ARP} and \ac{ICMP} packets.
This wait time between packets is set to a high value, because \ac{ARP} requests are broadcast to the complete broadcast domain and, as a result of this, it affects all devices within this subnet.
Further, the delay between each single port scan is also set to 100$ms$ to reduce the network load.
Both delays can be changed easily to fulfill the custom requirements of a certain industrial network.

At the begin, the scan is delayed for a random time of some seconds, that, after a power up,
not all edge nodes start scanning the same \ac{IP} at a time (see \Cref{subsec:pseudo_random_scanning}).
After this, \ac{ARP} requests for each \ac{IP} address are sent out, resulting in a maximum of 4 packets/s.
If an \ac{ARP} response is received, an \ac{ICMP} ping is executed (max 4 packets/s).
This means that the host is reachable and the first 1024 \ac{TCP} ports are scanned,
which is done with a maximum of about 25 packets per second.
This depends on the state of the port, e.g. if it is open or closed.
In comparison, the standard Modbus/TCP traffic in our testbed is about 400 packets per second between each node and the \ac{PLC}.
To distribute the load, if more nodes are scanning, the host selection is pseudo randomized (see \Cref{subsec:pseudo_random_scanning}).

An overview of packet sizes is given in \Cref{tab:packetbytes}.
For example, 25 \textit{SYN} packets/s with a size of 60 bytes each generate
a throughput of 1500 bytes/s (12 kbits/s).
The 25 packets/s is a mixed calculation for open and closed ports,
since, for example, with closed ports, only one \textit{ACK}+\textit{RST} returns from the target.
In contrast to that, an open port results in a three-way handshake as illustrated in \Cref{fig:scantype}.

\begin{table}[htb]
    \centering
    \caption{Data size of different packets from our scanner or as a response to it.}
    \label{tab:packetbytes}
        \begin{scriptsize}
        \begin{tabular}{l l}
            \hline
            \textbf{Packet}        & \textbf{Bytes} \\
            \hline
            \acs{ARP} request       & 60 \\
            \acs{ARP} reply         & 60 \\
            \acs{ICMP} ping request & 74 \\
            \acs{ICMP} ping reply   & 74 \\
            \acs{TCP} SYN           & 60 \\
            \acs{TCP} SYN/RST       & 60 \\
            \acs{TCP} SYN/ACK       & 60 \\
            \acs{TCP} FIN           & 60 \\
            Example SSH banner     & 95 \\
        \end{tabular}
        \end{scriptsize}
\end{table}

\subsection{\ac{MCU} Requirements}
\Cref{tab:build} shows the build output of the different sections in bytes.
The FreeRTOS task uses a maximum of 2048 words and can run with a low priority.
Additionally, the time between packets can be set to a high value, which results in a sleep (blocked state) of the scan task, whereby other operations can be performed.

\begin{table}[htb]
    \centering
    \caption{Binary comparison of example application with and without scanner.}
    \label{tab:build}
        \begin{tabular}{l l l l l l}
            \hline
            \textbf{Information} & \textbf{text} & \textbf{data} & \textbf{bss} & \textbf{dec} & \textbf{hex} \\
            \hline
            With scanner    & 140040 & 12588 & 293704 & 446332 & 6cf7c \\
            Without scanner & 129368 & 12588 & 293552 & 435508 & 6a534 \\
            \hline
            Difference      & 10672  & 0     & 152    & 10824  & 2a48  \\
        \end{tabular}
\end{table}

Most of the \acp{MCU} enabling networking should have enough performance to handle the additional scan task,
because of the relatively low RAM and ROM requirements.
Nevertheless, by optimizing the code,
the requirements of the scan building block can be further reduced.

\subsection{Attacker and Detection Consideration}
The detection in the testbed depends on the scenario and the configuration of the attacker device.
Furthermore, a trusted scan with a clean network at the beginning must be ensured.
For this reason, five possible attack scenarios are modeled:

\begin{enumerate}[label=\protect\circled{\color{white}\arabic*}]
    \item One edge node is removed from the network.
          This can happen, e.g., when an attacker removes the device or by a malfunction.
          The type of attack requires little knowledge of the specific target.
          Therefore, the attacker is considered weak.
    \item Services offered in the network have disappeared or new services have been added.
          This can happen when an adversary attacks services which crash
          or introduces back-doors that open new ports.
          This type of attack requires a moderate attacker knowledge, because changes to the network are made.
    \item An attacker attaches a standard configured computer to the network.
          There is no special configuration made by the attacker to be undetectable.
          Adding a computer to perform e.g. a port scan requires little knowledge,
          which can be done by a weak attacker.
    \item A \ac{MitM} attack is executed.
          In this case, the attacker has complete control over the traffic
          between two or more network participants.
          This enables viewing and manipulating the data.
          For this scenario, the attacker model is medium,
          because of the necessary high privileges.
    \item An attacker is performing a ``stealth'' attack \cite{singh2015network}.
          For example, the attacker is passively listening to the network traffic and makes a ``stealth'' port scan.
          This passive attack is a special attack on a network, where a system is secretly
          monitored and scanned e.g. for open ports and vulnerabilities.
          The purpose is solely to collect information about the network and hosts.
          No data is being injected into the destination network by the attacker.
          This scenarios requires a strong attacker model,
          because of the necessary knowledge.
\end{enumerate}

For scenario \circled{\color{white}1}, our network scanner detects the changes, because the host is not reachable by pings anymore and can handle the intrusion.
If services with open ports change (scenario \circled{\color{white}2}), they are detected by the port scan.
A standard configured computer (scenario \circled{\color{white}3}) even without open ports can be found by \ac{ICMP} pings.
If a \ac{MitM} attack (scenario \circled{\color{white}4}) is executed, in some cases the latency of pings is getting higher.
In this case, the \ac{MitM} attack could be detected by analyzing the ping timing, otherwise it is not possible with this approach.
Stealth attacks or passive listening (scenario \circled{\color{white}5}) cannot be detected by active scanning methods, like the here presented edge node scanner.
\Cref{tab:attacks_sumamry_detection} summarizes the detection of the different scenarios.

\vspace*{-0.3cm}
\begin{table}[htb]
    \centering
    \caption{Summary of the evaluated attack scenarios and detection capabilities.}
    \label{tab:attacks_sumamry_detection}
        \begin{tabular}{l l l l l}
            \hline
            \textbf{Model} & \textbf{Short description} & \textbf{Attacker} & \textbf{Detection} & \textbf{Mechanism} \\
            \hline
            \circled{\color{white}1}   & Node removed      & weak   & \cmark & \acs{ICMP} ping \\
            \circled{\color{white}2}   & Service changed   & medium & \cmark & SYN scan \\
            \circled{\color{white}3}   & Standard attack   & weak   & \cmark & \acs{ICMP} ping \\
            \circled{\color{white}4}   & \acs{MitM} attack & medium & \omark & timing \\
            \circled{\color{white}5}   & Stealth attack    & strong & \xmark & -- \\
            \hline
            \multicolumn{4}{r}{\cmark detected \omark dependent \xmark not detected}
        \end{tabular}
\end{table}

\subsection{``Stealth'' Attacker Configuration}
Case \circled{\color{white}5} is possible, if the attacker suppresses any network interaction from outside,
has no open ports, and disables \ac{ICMP} echo (\Cref{lst:disableicmp}).

\begin{lstlisting}[label={lst:disableicmp},
		           caption={Command to disable ARP and ICMP echo in Linux},
		           breaklines,
                   language=bash,
		           numbersep=3pt,
		           frame=lines,
                   framerule=0.5pt,
                   framexleftmargin=2em
                   ]
echo "1" >  /proc/sys/net/ipv4/icmp_echo_ignore_all
ip link set dev enp0s31f6 arp off
\end{lstlisting}

In this case, it is not possible for the network scanner to detect the device.
Though, the attacker must know,
if there is a continuous network scanning to get not detected.
Additionally, knowledge is required how the network is configured
and in local networks is not allowed to
response to \ac{ARP} requests.

However, the evaluation in our testbed has shown the feasibility of our approach
with a minimum of additional network load.
Furthermore, it is possible to link this security building block with other,
e.g. in combination with intrusion detection systems.

\section{Conclusion \label{sec:conclusion}}
\balance
In this paper, we introduced a network scanning and mapping building block for embedded low-cost \ac{IIoT} edge node devices.
We showed the feasibility of our approach in our open-source industrial testbed.
Our active scanner network mapping approach is lightweight and the results are clear and
detailed in contrast to most passive network monitoring approaches.

The amount of additional traffic in the network with our sample configuration with a mean of 4 packets/s
and peeks up to 25 packets/s from a single edge node
is low and could be adjusted if necessary.
Furthermore, the integration of our FreeRTOS scanning task and the configuration for the network can easily be done in other projects.
With our building block the security level of low-cost edge node devices,
e.g. for securing the \ac{IIoT}, can be increased.

\bibliographystyle{ieeetr}
\bibliography{\jobname}


\end{document}